\documentclass[final,5p,times,twocolumn]{elsarticle}
\usepackage{amssymb}
\usepackage{graphicx}
\usepackage{amsmath}
\usepackage{adjustbox}
\usepackage{gensymb}
\usepackage{amsfonts}
\usepackage{color}
\usepackage{amscd}
\usepackage{bm}

\journal{Chinese Physics C}
\makeatletter
\def\ps@pprintTitle{%
  \let\@oddhead\@empty
  \let\@evenhead\@empty
  \let\@oddfoot\@empty
  \let\@evenfoot\@empty}
\makeatother
\makeatletter
\def\ps@pprintTitle{%
  \let\@oddhead\@empty
  \let\@evenhead\@empty
  \let\@oddfoot\@empty
  \let\@evenfoot\@empty}
\makeatother

\begin{document}

\begin{frontmatter}

\title{Quantum gravitational corrections to Reissner-Nordström black hole thermodynamics and their implications for the weak gravity conjecture}


\author[1st,2nd,3rd]{Yong Xiao\corref{cor1}}
\author[1st,2nd]{Qiang Wang}
\author[1st,2nd]{Aonan Zhang}

\cortext[cor1]{Email:xiaoyong@hbu.edu.cn}

\affiliation[1st]{organization={Key Laboratory of High-precision Computation and Application of Quantum Field Theory of Hebei Province,
College of Physical Science and Technology, Hebei University},
            city={Baoding},
            postcode={071002},
            country={China}}
\affiliation[2nd]{organization={Hebei Research Center of the Basic Discipline for Computational Physics},
            city={Baoding},
            postcode={071002}, 
            country={China}}
\affiliation[3rd]{organization={Higgs Centre for Theoretical Physics, School of Mathematics, University of Edinburgh},
            city={Edinburgh},
            postcode={EH9 3FD}, 
            country={United Kingdom}}
        
\begin{abstract}
In this paper, we investigate the quantum gravitational corrections to the thermodynamical quantities of Reissner-Nordström black holes within the framework of effective field theory. The effective action originates from integrating out massless particles, including gravitons, at the one-loop level. We perform a complete thermodynamic analysis for both non-extremal and extremal black holes, and are mainly concerned about the shift in the charge-to-mass ratio $q/M$ that plays an important role in analyzing the weak gravity conjuecture. For non-extremal black holes, we identify a relationship between the shift in the charge-to-mass ratio and the thermodynamic stability of the black holes. For extremal black holes, we show that quantum gravity effects naturally lead to the super-extremality $q/M>1$  of charged black holes.
\end{abstract}



\end{frontmatter}




\section{Introduction}

In Einstein's general relativity, the extremality bound for a Reissner-Nordström (RN) black hole is given by $\frac{q}{M}\leq 1$ in natural units, where $q$ (implicitly $|q|$) and $M$ are the electric charge and mass of the black hole, respectively. On the other hand, for an extremal charged black hole, postulate that it has the ability to decay into two or more lighter components, then, due to the conservation of energy and charge, one of the decay products must have a charge-to-mass ratio $\frac{q}{M}> 1$. If such a super-extremal object doesn't exist in nature, the decay process would be completely forbidden, and the evolution of black holes would cease at a final stage. Eventually, the universe would be populated by a large number of stable remnants with various charges, which is an unfavorable scenario.

This situation has motivated the so-called weak gravity conjecture (WGC). The WGC suggests that for any $U(1)$ gauge field coupled to gravity, there must exist at least one object with charge $q$ and mass $M$ such that its charge-to-mass ratio exceeds the bound of a large extremal black hole \cite{Arkani-Hamed:2006emk}. As the name implies, consistent theories must include states where electromagnetic repulsion dominates gravitational attraction. Currently, there are various versions of the WGC, both mild and strong, with different motivations, interpretations, and implications. For an in-depth review of this field, see Ref. \cite{Harlow:2022ich}.

Proving the WGC is a challenging task with multiple approaches. A natural idea is that the well-known bound $q\leq M$ is a result of Einstein's gravity, which is a classical theory from a modern perspective. Thus, when quantum gravity effects are properly considered, the black hole may possess a state with $q > M$, allowing an extremal charged black hole to decay further.

In this regard, higher-derivative terms should be incorporated into the gravitational effective action as manifestations of quantum gravity effects \cite{Boulware:1985wk,Cardoso:2018ptl,BarrosoVarela:2023ull}. The quantum-corrected effective action can be formally written as 
\begin{align}
\begin{split}
I_{\text{tot}} &= \int_{\mathcal{M}} \frac{1}{16\pi}(R-F_{\mu\nu}F^{\mu\nu})+\epsilon L_{\text{hd}},
\end{split}\label{EFTaction}
\end{align}
where $L_{\text{hd}}$ represents the higher-derivative terms and the parameter $\epsilon$ is introduced to track their effects. Researchers can study the variations in the mass and entropy of an extremal RN black hole caused by turning on these terms. Typically, a decrease in the mass $M_{\text{ext}}$ at a fixed charge $q$ is expected. Therefore, proving the WGC is equivalent to verifying whether the following expression holds:
\begin{align}
\left(\frac{\partial M_{\text{ext}}(q,\epsilon)}{\partial \epsilon}\right)_q < 0.
\label{dmdsrelation}
\end{align}
Significant achievements have been gained in recent years \cite{Kats:2006xp,Cremonini:2009ih,Cremonini:2019wdk,Cheung:2018cwt,Cheung:2019cwi,Noumi:2022ybv}. Specifically, four-derivative terms such as $R^{2}$, $R_{\mu \nu}^{2}$ and $R_{\mu \nu \alpha \beta}^{2}$ have been extensively studied, and eq.\eqref{dmdsrelation} is utilized to place constraints on the combination of the coefficients of these curvature terms. There is a growing body of evidence in support of the WGC, coming from concrete models constructed from string theories or under certain assumptions. In these studies, local $\mathcal{O}(R^{2})$ terms are often considered the primary contributors to relevant corrections. However, loop effects of massless particles including gravitons can dominate over these $\mathcal{O}(R^{2})$ terms in the IR regime and provide the leading contributions \cite{Harlow:2022ich}. This underscores the importance and necessity of considering loop effects in the context of the WGC. Indeed, Refs.\cite{Charles:2019qqt,Arkani-Hamed:2021ajd} investigated loop effects of minimally coupled massless particles and found that the WGC is satisfied for extremal RN black holes.

Given the importance of this problem, we revisit the question from an alternative approach. We observe the existence of a non-local effective action that captures the loop effects of massless particles \cite{tHooft:1974to,Deser:1974em,Barvinsky:1985an,Donoghue:1994dn,Donoghue:2014yha,Calmet:2019eof,Calmet:2024neu,Teixeira:2020kew,qfcs}. Naturally, our motivation is to investigate whether such effects can reduce the mass of a RN black hole at fixed charge $q$. The advantage of our formalism is that it enables us to obtain all thermodynamic quantities for a quantum-corrected RN black hole. We find that the loop effects reduce the mass of RN black holes even far from their extremal limit, specifically, in the regime where the classical black hole parameters satisfy $q\leq \bar{r}_h< \sqrt{3}q$, where $\bar{r}_h$ denotes the corresponding outer horizon radius. This accommodates the results of Refs.\cite{Charles:2019qqt,Arkani-Hamed:2021ajd} for extremal RN black holes as a special case  $ \bar{r}_h \to q$ of our findings. Notably, $q\leq\bar{r}_h <\sqrt{3}q$ also corresponds to the parameter range for thermodynamically stable RN black holes. In addition, our complete thermodynamic analysis interestingly shows that the loop effects simultaneously generate both the super-extremality required by the WGC and the logarithmic entropy corrections, which were previously considered as separate research themes.

The structure of the remaining paper is as follows. In Sec. \ref{sec2}, we present the effective action that describes one-loop corrections to the Einstein-Maxwell theory. In Sec. \ref{sec3}, we compute the quantum corrections to the Euclidean action and free energy of the RN black hole induced by one-loop effects. The subsections there are dedicated to analyzing the thermodynamic corrections for non-extremal and extremal RN black holes respectively. Sec. \ref{sec4} contains the concluding remarks. Additionally, the Appendix includes a comparison between our effective action and those employed in other related works \cite{Charles:2019qqt,Arkani-Hamed:2021ajd}. For the convenience of interested readers seeking to reproduce our results, all the detailed computations are available in the auxiliary notebook of the arXiv version of this paper.

\section{One-Loop Effective Action of Quantum Gravity} \label{sec2}
Thanks to the pioneering works \cite{tHooft:1974to,Deser:1974em,Barvinsky:1985an,Donoghue:1994dn,Donoghue:2014yha,Calmet:2019eof,Calmet:2024neu,Teixeira:2020kew,qfcs}, the quantum corrections to Einstein--Maxwell theory can be organized systematically in an effective field theory expansion. In this framework, one treats the classical metric and gauge field as background fields, integrates out the quantum fluctuations (in particular, the massless modes such as gravitons) in the background-field formalism, and then expands the resulting one-loop effective action in powers of curvatures and derivatives. A key output of this construction is that, besides the local counterterms that renormalize the curvature-squared operators, massless loops unavoidably generate genuinely nonlocal form factors. To the leading  order in the curvature expansion, the higher-derivative part can be written as
\begin{align}
\begin{split}
& L_{\text{hd}} = \left[ c_1(\mu) R^2 + c_2(\mu) R_{\mu\nu} R^{\mu\nu} + c_3(\mu) R_{\mu\nu\rho\sigma} R^{\mu\nu\rho\sigma} \right] \\
&\!-\!\left[\alpha R \ln(\frac{\Box}{\mu^2} )R \!+\! \beta R_{\mu\nu} \ln(\frac{\Box}{\mu^2}) R^{\mu\nu} \!+\! \gamma R_{\mu\nu\alpha\beta} \ln( \frac{\Box}{\mu^2} ) R^{\mu\nu\alpha\beta} \right]\!.
\end{split}\label{EFTaction1}
\end{align}
The physical foundation of eq.~\eqref{EFTaction1} rests on the standard low-energy effective description of quantum gravity: the local coefficients $c_i(\mu)$ encode UV-sensitive matching information and depend on the renormalization prescription, while loops of massless fields induce universal, long-range nonlocalities encoded in operators such as $\ln\left(\frac{\Box}{\mu^2}\right)$, where $\Box$ denotes the d'Alembert operator. The computation of one-loop divergences (and hence the relevant Wilson coefficients $\alpha$, $\beta$, and $\gamma$) dates back to the 1970s \cite{tHooft:1974to,Deser:1974em}. The modern effective-field-theory perspective, together with its predictive nonlocal sector, was formalized and systematized in Donoghue's seminal works \cite{Donoghue:1994dn}. Recently, this framework continues to attract interest and has been applied to quantum gravity phenomenology and black hole physics\footnote{The study of such nonlocal effects on black hole solutions and other semiclassical corrections to gravity can be traced back to Refs. \cite{Duff:1974ud,Duff:1977ay}.} (see, e.g., \cite{Calmet:2019eof,Calmet:2021stu,Boasso:2025ptx,Bukhari:2025fhd}). In particular, in cosmological scenarios with a time-dependent background, it has been shown how the effective action yields causal behaviors and manifests nonlocality as a memory effect \cite{Donoghue:2014yha}.

By construction, the effective action is renormalization group invariant, namely the explicit $\mu$-dependence of $c_i(\mu)$ cancels against that of $\ln\left(\frac{\Box}{\mu^2}\right)$ \cite{El-Menoufi:2015cqw}. For later convenience, we will absorb $c_i(\mu)$ into its partner $\ln\left(\frac{\Box}{\mu^2}\right)$, thereby defining the operator $\ln\left(\frac{\Box}{\mu_{*}^2}\right)$. Here, $\mu_{*}$ represents the characteristic energy scale beyond which the effective field-theoretical description breaks down; it should be fixed by the UV completion of quantum gravity \cite{Calmet:2024neu}. Without loss of generality, we take $\mu_{*}$ to be near the Planck scale. Throughout this paper we work in the regime $r \gg 1/\mu_*$, where the effective field theory expansion is under control.

We are particularly interested in this effective action for two main reasons. Firstly, the Wilson coefficients $\alpha$, $\beta$, and $\gamma$ of the nonlocal terms are calculable and insensitive to concrete UV theories. This insensitivity represents a model-independent prediction of the effective theory. Since the RN black hole is Ricci-flat, the Wilson coefficient $\alpha$ does not contribute. Below we list the values of $\beta$ and $\gamma$:
\begin{align}
    \beta &= \frac{1}{11520\pi^2}(-2 n_s + 8 n_f + 176 n_v +52), \label{a2}\\
    \gamma &= \frac{1}{11520\pi^2} (2n_s + 7 n_f - 26 n_v +398 ), \label{gamma}
\end{align}
where $n_s$, $n_f$, $n_v$ denote the numbers of minimally coupled scalars, four-component fermions, additional vectors other than the background $U(1)$ gauge field in the low-energy particle spectrum of nature. The contributions from the fluctuations of the gravitational and Maxwell fields mix together and combine to produce the two numbers $52$ and $398$. Note that these numerical coefficients represent one-loop corrections to the Einstein-Maxwell theory (see Refs. \cite{Charles:2015eha,Bhattacharyya:2012wz} for details of the derivation). This explains why they differ slightly from those provided in \cite{Donoghue:2014yha,Calmet:2019eof,Calmet:2024neu}, which correspond to one-loop corrections to Einstein theory.  

In principle, in Einstein--Maxwell theory, the effective action \eqref{EFTaction1} may also contain operators involving the Maxwell field strength, such as $R_{\mu\nu\rho\sigma}F^{\mu\nu}F^{\rho\sigma}$ and $F^4$. However, using the background-field equations of motion and Bianchi identities, these terms can be eliminated, yielding a purely geometric result \cite{Deser:1974em,Bhattacharyya:2012wz,Charles:2017dbr}. Conversely, such terms can be reintroduced by field redefinitions, which also alter the value of $\beta$. Thus, the coefficients in eq.\eqref{a2} should be understood as those in a gauge where the Maxwell sector has no logarithmic running coefficients at the one-loop level.

Secondly, the nonlocal terms are at the leading order in the curvature expansions, which makes their effects the most prominent. Specifically, even if one takes into account the additional local curvature terms $L_{\text{hd}}=c_2 R^2+c_3 R^3+c_4 R^4+ \cdots $ that are constructed from a UV theory in a top-down manner, their contributions are overshadowed by those of the $R\ln\Box R$ terms. According to the scaling behaviors of these terms, their contributions can be arranged in a sequence: $\ln(r^2\mu_{*}^2)$, $1$, $\frac{1}{r^2}$, $\frac{1}{r^4}$, $\cdots$. When working in the regime $r > 1/\mu_{*}$, the first term dominates over all other contributions. In contrast, when approaching the scale $r\sim 1/\mu_{*}$, all contributions are expected to be of the same order, and the effective field theory becomes invalid. Only a complete quantum gravity theory can sum over all these contributions. Since we are working within the regime $r > 1/\mu_{*}$, we will focus on the nonlocal terms and analyze their implications for the WGC, while ignoring all the contributions from local terms.

\section{Thermodynamic Corrections of the RN Black Hole} \label{sec3}
To prove eq.\eqref{dmdsrelation}, we need to calculate the corrections to the thermodynamic quantities of a RN black hole, such as its mass and entropy, which are induced by switching on $\epsilon L_{\text{hd}}$.

Conventionally, the expressions of thermodynamic quantities should be derived from the black hole metric using standard techniques. At the first order in $\epsilon$, the perturbed RN black hole metric can be formally written as
\begin{align}
    g_{\mu \nu}=\bar{g}_{\mu \nu}+\epsilon g_{\mu \nu}^{(1)},
\end{align}
 where $\bar{g}_{\mu \nu}$ represents the original unperturbed RN metric $ds^2 = -f(r)dt^2 + f^{-1}(r)dr^2 + r^2d\Omega^2$, and $f(r) = 1-\frac{2M}{r}+\frac{q^2}{r^2}$. However, determining the exact form of the perturbed metric induced by the nonlocal terms in eq.\eqref{EFTaction1} is far from an easy task. Due to the lack of appropriate techniques to handle the variation $\delta \ln \Box$, the gravitational field equations cannot be derived straightforwardly from the variation analysis of the Lagrangian. Thus, certain assumptions have to be made to obtain the perturbed field equations and the corresponding black hole solution. This has been done in our previous work \cite{Xiao:2021zly}.

Anyway, next we will adopt another method to dealing with this problem independent of the techniques used in Ref.\cite{Xiao:2021zly}\footnote{We simply use Ref.\cite{Xiao:2021zly} for the purpose of cross-checking of the results.}. We notice that there exists a convenient approach indicating that, at the first order in $\epsilon$, the corrections to the black hole's thermodynamic quantities from any higher-derivative terms $\epsilon L_{\text{hd}}$ can be directly obtained using the \emph{unperturbed} metric $\bar{g}_{\mu \nu}$, without knowing the exact form of $g_{\mu\nu}=\bar{g}_{\mu\nu}+\epsilon g_{\mu \nu}^{(1)}$. This approach was discovered occasionally by Guber et al. \cite{Gubser:1998nz} and rediscovered and well-explained recently by Reall and Santos \cite{Reall:2019sah}. For more discussions on its validity and recent applications, refer to Refs.\cite{Caldarelli:1999ar,Xiao:2022auy,Xiao:2023two,Ma:2023qqj,Hu:2023gru,Ma:2024ynp,Guo:2025muo,Chen:2025ary}. By following this approach, we can deduce the expression of $M_{ext}(q,\epsilon)$ without the need of making unnecessary assumptions.

We now start to evaluate the onshell Euclidean action 
\begin{align}
    I_{\text{tot}}(g_{\mu \nu}) = I_{\text{RN}}+\epsilon I_{\text{hd}}, \label{Icalculate}
\end{align}
by substituting $g_{\mu \nu}$ into it. The Euclidean action is directly related to the partition function and free energy of the system. According to the aforementioned approach, when evaluating the Euclidean action, using the perturbed metric $g_{\mu \nu}$ or the unperturbed metric $\bar{g}_{\mu \nu}$ can yield the same result at the first order in $\epsilon$. The reasoning is as follows. 

The onshell Euclidean action \eqref{Icalculate} comprises two components: $I_{\text{RN}}$ and $\epsilon I_{\text{hd}}$, which can be evaluated separately. First, consider the two-derivative term $I_{\text{RN}} \equiv \frac{1}{16\pi} \int_{M} (R - F_{\mu\nu}F^{\mu\nu})$ along with its boundary term. Since the original RN metric $\bar{g}_{\mu\nu}$ corresponds to the extremum of this functional integral, any perturbation around $\bar{g}_{\mu\nu}$ will only induce a correction of $\mathcal{O}(\epsilon^2)$. This implies that $I_{\text{RN}}(g_{\mu\nu}) = I_{\text{RN}}(\bar{g}_{\mu\nu})$ at $\mathcal{O}(\epsilon)$, provided that $g_{\mu \nu}$ and $\bar{g}_{\mu \nu}$ obey the same boundary conditions\footnote{In functional analysis, a meaningful comparison between $\mathcal{F}[f_1(x)]$ and $\mathcal{F}[f_2(x)]$ requires the functions $f_1(x)$ and $f_2(x)$ to obey the same boundary conditions. For the present application, this means the perturbed metric $g_{\mu\nu}$ and unperturbed metric $\bar{g}_{\mu\nu}$ must share the same temperature and charge.}.  This analysis also holds when changing the thermodynamic ensemble \cite{Cremonini:2019wdk}. 

As a result, when evaluating the first part $I_{\text{RN}}(g_{\mu\nu})$ of the Euclidean action \eqref{Icalculate}, one can directly adopt the well-known textbook expression for the Euclidean action $I_{\text{RN}}(\bar{g}_{\mu\nu})$ of the original RN black hole. In an ensemble characterized by temperature $T$ and charge $q$, the corresponding free energy is given by  
\begin{align} 
F_{\text{RN}}(T,q) = \frac{3 q^2 + \bar{r}_h^2}{4 \bar{r}_h}, \label{free0} 
\end{align}  
which is directly borrowed from the known formula for the unperturbed RN black hole with the same temperature $T$ and charge $q$. Note that passing from the $(T,\Phi)$ ensemble to the $(T,q)$ ensemble can be strictly realized by a Legendre transformation, i.e., adding a total derivative term $\frac{1}{4\pi}\int_\mathcal{M}\nabla_\mu(F^{\mu\nu}A_\nu)$ to the action \eqref{EFTaction}.

In eq.\eqref{free0}, $\bar{r}_h$ represents the outer horizon radius of the unperturbed black hole, and the expression of $\bar{r}_h(T,q)$ should be solved from the relation \begin{align} T=\frac{1}{4 \pi \bar{r}_h}-\frac{q^2}{4 \pi \bar{r}_h^3}, \label{temrh} \end{align} and then inserted into eq.\eqref{free0}. In practice, this is not necessary because we can use the chain rule to compute quantities such as $\frac{\partial F }{\partial T}$. Technically, we could also re-express eq.\eqref{free0} as a function of the horizon radius $r_h=\bar{r}_h+\mathcal{O}(\epsilon)$ of the perturbed black hole, but it is more convenient to use $\bar{r}_h$. For example, taking the external limit $T\rightarrow 0$ is equivalent to taking $\bar{r}_h\rightarrow q$. If we use $r_h$ as the variable, all such expressions will soon become complicated. For those who are confused in thinking back and forth between the characteristic quantities of a perturbed and unperturbed black hole, it is helpful to ignore the origin of $\bar{r}_h$ and simply regard it as a convenient parameter related to the temperature $T$ and charge $q$ through eq.\eqref{temrh}. Then, one can focus on the perturbed black hole that we are actually concerned with. By the way, another intriguing observation from eq.\eqref{free0} is that the expression of $F_{RN}(T,q)$ indeed receives corrections at $\mathcal{O}(\epsilon)$ using the horizon radius $r_h$ as the variable, but such corrections are cleverly absorbed into the parameter $\bar{r}_h$.

Next, consider the second part $\epsilon I_{\text{hd}}\equiv\epsilon \int_\mathcal{M} L_{\text{hd}}$. It is obvious that $\epsilon I_{\text{hd}}(g_{\mu \nu})=\epsilon I_{\text{hd}}(\bar{g}_{\mu \nu})+\mathcal{O}(\epsilon^2)$ due to the presence of the factor $\epsilon$. Consequently, the unperturbed RN metric $\bar{g}_{\mu\nu}$ can be directly employed for the calculation, since we only need the results at $\mathcal{O}(\epsilon)$. The other problem is that the calculation of the terms like $\ln(\frac{\Box}{\mu_{*}^2}) R_{\mu \nu \alpha \beta}$ would be tricky. However, progresses have been made in Ref.\cite{Calmet:2019eof,Delgado:2022pcc,El-Menoufi:2017kew}, where explicit calculations show that its main contribution has the form $-\ln (r^2 \mu_*^2)  R_{\mu \nu \alpha \beta}$, after appropriate regularization procedure. Substituting it into $\epsilon \int_{\mathcal{M}} L_{hd}$ and calculating the integral from the horizon to infinity, we can obtain the expression of $\epsilon I_{\text{hd}}(\bar{g}_{\mu \nu})$. Besides, for the asymptotical-flat cases, the boundary terms associated with the higher derivative part of the action decay rapidly as $r\rightarrow \infty$, therefore can be neglected in the evaluation of $I_{\text{hd}}$ \cite{Reall:2019sah}.

In total, at the first order in $\epsilon$, the free energy of the perturbed RN black hole is obtained as
\begin{align}
\begin{adjustbox}{width=\linewidth-2.4em}
$ \displaystyle
F(T,q)  \!=\!  \frac{3 q^2 \!+\! \bar{r}_h^2}{4 \bar{r}_h}
\!-\! \frac{16 \pi  \epsilon  \left(   (\beta \!+\! 4 \gamma )q^4 \!-\! 5 \gamma  q^2 {\bar{r}_h}^2 \!+\!   5 \gamma  {\bar{r}_h}^4\right)\ln \left( {\bar{r}_h}^2\mu_* ^2\right) }{5 {\bar{r}_h}^5}.
$
\end{adjustbox}\label{freetqPd}
\end{align}
To highlight the universal logarithmic corrections, we have omitted the subleading terms, as $\ln(\bar r_h^2\mu_*^2 ) \gg 1$. These subleading terms are also non-universal because their values can be easily altered by other model-dependent local curvature-squared interactions. 

\subsection{The non-extremal RN black hole and the shift of charge-to-mass ratio}

All thermodynamic quantities can be derived from the free energy $F(T,q)$ of the ensemble depending on $T$ and $q$. The entropy and eletrostatic potential of the system can be calculated from $S=-\frac{\partial F}{\partial T}$ and $\Phi=\frac{\partial F}{\partial q}$, and the mass can be calculated from $M=F+TS$. The results are consistent with our previous work \cite{Xiao:2021zly}, where we explicitly solved the perturbed black hole metric under some reasonable assumptions. 

Specifically, the mass and entropy are respectively obtained as
\begin{align}
\begin{adjustbox}{width=\linewidth-2.92em}
$ \displaystyle
    M\!=\! \frac{q^2 \!+\! \bar{r}_h^2}{2 \bar{r}_h}\!+\! \frac{32 \pi \epsilon q^4   (\beta \!+ \! 4 \gamma ) \left(  2 \bar{r}_h^2  \!-\! q^2\right) \ln \left( \bar{r}_h^2\mu_* ^2\right)}{5 \bar{r}_h^5 \left(\bar{r}_h^2-3 q^2\right)},
$
\end{adjustbox}\label{massnonE}
\end{align}
and
\begin{align}
\begin{adjustbox}{width=\linewidth-2.86em}
$ \displaystyle
S=\pi  \bar{r}_h^2 \!+\! \frac{64 \pi ^2 \epsilon \left((\beta \!+\! 4 \gamma )q^4 \!-\! 3 \gamma  q^2 \bar{r}_h^2\!+\!\gamma  \bar{r}_h^4\right)  \ln \left( \bar{r}_h^2\mu_* ^2\right) }{\bar{r}_h^2 \left(\bar{r}_h^2-3 q^2\right)}.
$
\end{adjustbox}\label{entrgen}
\end{align}
The eletrostatic potential at the horizon is 
\begin{align}
\Phi=\frac{q}{\bar{r}_h} +  \frac{32 \pi \epsilon     (\beta +4 \gamma ) \left(q^2-2 \bar{r}_h^2\right)q^3 \ln \left( \bar{r}_h^2\mu_* ^2\right)}{5 \bar{r}_h^5 \left(\bar{r}_h^2-3 q^2\right)}. \label{qphi}
\end{align}
The first law of RN black hole $dM=TdS+\Phi dq$ automatically holds.

The Smarr relation can be derived from the first law via a scaling argument. In the present case, the effective action involves the dimensionful parameter $\epsilon$ and the characteristic energy scale $\mu_*$, so the Smarr relation must be extended by including their associated conjugate variables \cite{Kastor:2009wy,Xiao:2023lap}. Using the homogeneity property
\begin{align}
\eta\, M(S,q,\epsilon,\mu_*)=M(\eta^2 S,\eta q,\eta^2 \epsilon,\eta^{-1}\mu_*),
\end{align}
we obtain the Smarr relation for the quantum-corrected black hole:
\begin{align}
M = 2TS + q\Phi + 2 V_{\epsilon}\,\epsilon - V_{\mu_*}\,\mu_*, \label{smarrsub}
\end{align}
where $V_{\epsilon}$ and $V_{\mu_*}$ are computed from the free energy as $V_{\epsilon}=(\partial F/\partial \epsilon)_{T,q,\mu_*}$ and $V_{\mu_*}=(\partial F/\partial \mu_*)_{T,q,\epsilon}$. One can verify that this Smarr relation is indeed satisfied by the thermodynamic quantities derived above.

We are now ready to explore the quantum gravitational effects on RN black holes, with a particular focus on how their charge-to-mass ratio changes \cite{Solomon:2020pja}. In this context, we find that
\begin{align}
\left.\frac{\partial M}{\partial \epsilon}\right|_{T,q} = -\frac{\mathcal{N}}{300 \pi} \, q^4 \left(2 \bar{r}_h^2 - q^2\right) \frac{\ln\left(\bar{r}_h^2 \mu_*^2\right)}{\bar{r}_h^5 \left(3 q^2 - \bar{r}_h^2\right)}, \label{stability}
\end{align}
where $\mathcal{N}$ is defined below in eq. \eqref{combpart} and $\mathcal{N} > 0$ always holds. The singularity at $\bar{r}_h = \sqrt{3}q$ indicates a critical point, which coincides with the boundary between thermodynamically stable and unstable RN black holes as determined by the sign of the heat capacity $C_q$. Correspondingly, for a thermodynamically stable black hole within the parameter range $q \leq \bar{r}_h < \sqrt{3}q$, the ratio $q/M$ increases (since $\partial M/\partial \epsilon < 0$ at fixed $q$). In contrast, for an unstable black hole within $\bar{r}_h > \sqrt{3}q$, $q/M$ decreases. This reveals a correlation between the thermodynamic stability of RN black holes and the direction of the shift in their charge-to-mass ratio.

\subsection{Extremal RN black hole and the WGC}

Finally, we consider the special case of extremal RN black hole. By taking the extremal limit $T \rightarrow 0$, which is equivalent to $\bar{r}_h \rightarrow q$, we get
\begin{align}
\begin{adjustbox}{width=\linewidth-3.5em}
$ \displaystyle
    M_{ext} = \bar{r}_h-\frac{\epsilon   (n_s\!+\! 6 n_f \! +\! 12n_v\!+\! 274)\ln \left( \bar{r}_h^2\mu_* ^2\right)}{600 \pi  \bar{r}_h} ,
$
\end{adjustbox}\label{smarrreg}
\end{align}
and 
\begin{align}
\begin{adjustbox}{width=\linewidth-2.79em}
$ \displaystyle
    S_{ext}=\pi  \bar{r}_h^2\!-\!\frac{\epsilon}{180}   (n_s\!+\! 11n_f \!+\! 62n_v \! +\! 424 ) \ln \left( \bar{r}_h^2\mu_* ^2\right). 
$
\end{adjustbox}\label{entrext}
\end{align}
In these equations, we have substituted the values of the Wilson coefficients $\beta$, and $\gamma$ from eqs.\eqref{a2} and \eqref{gamma}. As a consistency check, eq. \eqref{entrext} correctly reproduces the logarithmic correction to black hole entropy derived via an independent approach by Sen in \cite{Sen:2012kpz}, up to an additional term $-3 \ln A$ arising from the zero-modes associated with the rotational isometries of the extremal black hole \cite{Bhattacharyya:2012wz}.

The logarithmic entropy correction in our analysis stems from the same physical mechanism as in Sen's approach  \cite{Bhattacharyya:2012wz}, namely the one-loop effects of massless fields propagating in the black hole background. Within the effective field theory framework, these loop effects are encoded in nonlocal form factors (such as $\ln\Box$), giving rise to universal logarithmic corrections whose coefficients are determined by the low-energy particle spectrum; see also the related explicit one-loop computations for various black holes in \cite{Bhattacharyya:2012wz,Delgado:2022pcc,El-Menoufi:2017kew,Sen:2012dw,Sen:2011fc}.

On the other hand, logarithmic corrections to black hole entropy also emerge in other contexts with distinct physical origins. For instance, such terms arise from microscopic state counting in loop quantum gravity \cite{Kaul:2000rk}, from Cardy formula methods in conformal field theory \cite{Carlip:2000nv}, and from thermal and statistical fluctuations about the equilibrium state \cite{Das:2001ic}. While these results share a common $\ln (r_h^2)$ functional form, their coefficients and validity regimes are generally model- and ensemble-dependent, and do not necessarily coincide with the loop-induced coefficients derived from the low-energy effective action.

Now, from the expression of $M_{ext}$, we can straightforwardly obtain the result
\begin{align}
\left.\frac{dM_{\text{ext}}}{d\epsilon}\right|_q = -\frac{\mathcal{N}}{600\pi \bar{r}_h}\ln(\bar{r}_h^2\mu_*^2), \label{wgcpf}
\end{align}
where 
\begin{align}
    \mathcal{N} \equiv  n_s+ 6n_f  +  12n_v+274.\label{combpart}
\end{align}
Intriguingly, the coefficients conspire to produce a definite sign: since $n_s$, $n_f$ and $n_v$ represent the numbers of particles in the spectrum, it is certain that $\mathcal{N}>0$. Consequently, the loop effects of quantum gravity surely reduce the mass of an extremal charged black hole at a fixed charge $q$, which supports the analyses presented in Refs.\cite{Charles:2019qqt, Arkani-Hamed:2021ajd}. A more detailed comparison is provided in the Appendix.

The coefficient $\mathcal{N}$ does not depend on the UV theory of quantum gravity, whereas the characteristic energy scale $\mu_*$ does \cite{Calmet:2024neu}. Thus, the presence of $\ln\mu_*$ in eq.\eqref{wgcpf} indicates the UV sensitivity of our result. However, the model-dependent value of $\mu_*$ does not affect our conclusion for the WGC, since we work in the regime $r>1/\mu_*$ where the effective field-theoretical description is valid. Physics near the scale $r\sim 1/\mu_*$ would be described by a complete theory of quantum gravity and is not relevant to our current analysis.

Below we make further discussions on the computation of $\frac{\partial M_{ext}}{\partial \epsilon}$. While the above calculation is straightforward, it is worth noting that eq.\eqref{wgcpf} can also be derived from other formulas, which offer certain advantages. Concretely, we may utilize a crucial formula, proposed and proven in \cite{Goon:2019faz}, given by
\begin{align}
      \left(\frac{\partial M_{\text{ext}}(q,\epsilon)}{\partial \epsilon}\right)_q \!=\! -\! \lim\limits_{M\rightarrow M_{\text{ext}}}T\left(\frac{\partial S(M,q,\epsilon)}{\partial \epsilon}\right)_{M,q}.  \label{equalityMs}
\end{align}
In fact, from the extended first law $dM=TdS+\Phi dq +V_\epsilon d\epsilon$, it is straightforward to see that both sides of eq.\eqref{equalityMs} equal $\lim\limits_{T\rightarrow 0}V_\epsilon$. Then, from the relation $dF=-SdT+\Phi dq+V_\epsilon d\epsilon$, evaluating $\lim\limits_{T\rightarrow 0}V_\epsilon$ amounts to computing $\lim\limits_{T\rightarrow 0}\left(\frac{\partial F}{\partial \epsilon }\right)_{T,q}$, where $F=F_{RN}(T,q)-\epsilon \int d^3 x \sqrt{g} L_{\text{hd}}$. Therefore, $\frac{\partial M_{\text{ext}}}{\partial \epsilon}$ can be evaluated as the spatial integral $-\int d^3 x \sqrt{g} L_{\text{hd}}$ for an extremal RN black hole \cite{Cheung:2018cwt,Cheung:2019cwi}. This provides a quick way to check whether the WGC is satisfied for a given higher-derivative term $\epsilon L_{\text{hd}}$.

According to the formula \eqref{equalityMs}, another way towards the study of the WGC is to evaluate the shift in entropy due to switching on the quantum gravity effects. The key significance of this formula lies in its ability to link the decrease in mass to the increase in entropy. Notably, in this regard, our analysis establishes a novel connection between the required super-extremality and the logarithmic entropy corrections, i.e., both of eqs.\eqref{smarrreg} and \eqref{entrext} are generated by the same quantum gravity effects. A caveat should be noted here: the derivative $T\left(\frac{\partial S}{\partial \epsilon}\right)_{M,q}$ is evaluated at a fixed $M$ rather than at a fixed $T$. When $\epsilon$ is switched on while keeping $M$ constant, the temperature will be shifted away from its extremal value. For this reason, we should use eq.\eqref{entrgen} (applicable to general RN black holes) rather than the more familiar eq.\eqref{entrext} (which applies exclusively to extremal cases) for the calculation. It requires to use chain rule to execute the derivative and arrive at $T\left(\frac{\partial S}{\partial \epsilon}\right)_{M,q}>0$. This explains why previously no clues regarding the WGC have been derived from the logarithmic entropy corrections, despite such entropy corrections being known in the literature for many years.

\section{Concluding remarks} \label{sec4}
 
In this paper, we have investigated the effective action obtained from integrating out massless particles at the one-loop level. Our results demonstrate that the loop effects of quantum gravity can naturally induce super-extremality (\(q > M\)) in charged black holes. This suggests that quantum gravity plays a crucial role in facilitating the decay of extremal black holes into more fundamental components. These effects thus have far-reaching implications for our understanding of the ultimate evolution of black holes and the universe. We emphasize that the results are model-independent, as they depend solely on the low-energy particle spectrum of nature and are insensitive to the UV details of quantum gravity theories.

Note that our new findings are mainly encapsulated in eqs.\eqref{massnonE}--\eqref{entrext}.
Concretely, eqs.\eqref{massnonE}--\eqref{smarrsub} provide the complete thermodynamic behaviors of the RN black hole with one-loop corrections. Regarding eq.\eqref{stability}, it establishes a relationship between the shift in the charge-to-mass ratio and the thermodynamic stability properties of a general RN black hole\footnote{For context, a discussion on the correlation between thermodynamic stability and super-extremality can be found in Ref.\cite{Cheung:2018cwt}; however, this discussion was formalized based on tree-level higher-derivative terms for extremal black holes. Our results imply that this correlation can also be extended to one-loop effects and non-extremal cases.}. Furthermore, as the extremal limit of eqs.\eqref{massnonE} and \eqref{entrgen}, eqs.\eqref{smarrreg} and \eqref{entrext} reveal an unexpected connection between the super-extremality required by the WGC (reflected in the second term of $M_{ext}$) and logarithmic entropy corrections (reflected in the second term of $S_{ext}$), which were previously treated as separate research directions.

In summary, in the spirit of the swampland conjecture, any candidate for a consistent quantum gravity theory must satisfy certain fundamental physical criteria. As a pivotal principle of the swampland program, the WGC distinguishes consistent quantum gravity theories from those in the swampland. As noted in Ref.\cite{Solomon:2020pja}, the nonlocal features of loop processes remain far from well understood in the context of gravitational phenomena. Our work clarifies how the inclusion of loop effects establishes an intrinsic connection between quantum-corrected black hole thermodynamics and the WGC. Specifically, loop effects give rise to an entropy increase for a black hole with fixed mass $M$ and charge $q$. This behavior is physically intuitive: loop effects can blur the observational precision of the system's detailed properties \cite{Cheung:2018cwt,Reall:2019sah}. Then, according to eq.\eqref{equalityMs}, this entropy increase manifests as a corresponding mass increase. This analysis thus demonstrates that the WGC plays a fundamental role by being inherently embedded in the underlying operational framework of quantum gravity.

\section*{Acknowledgments}
We thank Grant Remmen, Nima Arkani-Hamed, and Yu-Tin Huang for their constructive discussions regarding loop effects of massless particles on WGC, which took place after our work was posted on arXiv. XY would like to thank Yu Tian and Hongbao Zhang for useful discussions. YX is also thankful to the Higgs Centre for Theoretical Physics at the University of Edinburgh for providing research facilities and hospitality during the visit. This work was supported in part by the National Natural Science Foundation of China with Grant No. 12475048, the Hebei Natural Science Foundation with Grant No. A2024201012, and the China Scholarship Council with Grant No. 202408130101.

\section*{Appendix: A comparison of effective actions}

Now we compare the effective action employed in our study with those adopted in previous works \cite{Charles:2019qqt, Arkani-Hamed:2021ajd}. First and foremost, the key distinction lies in the fact that our effective action is suited to consistently analyzing one-loop corrections to general RN black holes, whereas the actions in \cite{Charles:2019qqt, Arkani-Hamed:2021ajd} are constructed exclusively for the extremal case. In this Appendix, we demonstrate that our effective action is indispensable even for the extremal regime, as it is a prerequisite for establishing a self-consistent thermodynamic analysis.

We can adopt an alternative set of bases and rewrite eq.\eqref{EFTaction1} in the form:
\begin{align}
\begin{split}
& L_{\text{hd}} \!=\!  \!-\! \lambda_1 C_{\mu\nu\alpha\beta} \ln\left(\frac{\Box}{\mu_*^2}\right) C^{\mu\nu\alpha\beta} \!-\! \lambda_2 \left[ R \ln\left(\frac{\Box}{\mu_*^2}\right) R \right.  \\
&  \!-\! 4 R_{\mu\nu} \ln\left(\frac{\Box}{\mu_*^2}\right) R^{\mu\nu} \left. + R_{\mu\nu\alpha\beta} \ln\left(\frac{\Box}{\mu_*^2}\right) R^{\mu\nu\alpha\beta} \right],
\end{split}\label{EFTaction2}
\end{align}
where $C_{\mu\nu\alpha\beta}$ represents the Weyl tensor, and the second term exhibits a structure similar to the Gauss-Bonnet term. Given that the RN black hole is Ricci-flat, two independent bases of the higher-derivative terms are sufficient in this context. Using these bases, the combinations of coefficients $\lambda_{1}=\frac{1}{2}(\beta + 4\gamma)$ and $\lambda_{2}=-\frac{1}{2}(\beta+2\gamma)$ yield the values
\begin{align}
    \lambda_1 &= \frac{1}{3840\pi^2}( n_s + 6 n_f + 12 n_v +274), \label{a22}\\
    \lambda_2 &= -\frac{1}{11520\pi^2} (n_s + 11 n_f+62 n_v +424 ).
    \label{gamma2}
\end{align}
Interestingly, according to eqs.\eqref{entrext} and \eqref{combpart}, we can observe that $\lambda_{1}$ governs the behavior of super-extremality, while $\lambda_{2}$ controls the logarithmic corrections to the entropy of an extremal RN black hole.

By comparison, the higher-derivative part of the effective action in Refs.\cite{Charles:2019qqt,Arkani-Hamed:2021ajd} is often expressed in terms of four-photon interactions $F^{4}$. However, we note that, through an appropriate field redefinition, it can also be transformed into the form
\begin{align}
    L_{\text{hd}}=c_1 \,C_{\mu\nu\alpha\beta}C^{\mu\nu\alpha\beta} + c_2\, E_4 , \label{ceaction}
\end{align}
where $E_{4}$ is the standard Gauss-Bonnet term $R^{2}-4R_{\mu\nu}R^{\mu\nu}+R_{\mu\nu\alpha\beta}R^{\mu\nu\alpha\beta}$. The coefficients have a logarithmic running behavior with $c_{1}\sim\lambda_{1}\ln(\bar{r}_{h}^{2}\Lambda^{2})$ and $c_{2}\sim\lambda_{2}\ln(\bar{r}_{h}^{2}\Lambda^{2})$, where $\Lambda$ is the UV cutoff.

Apparently, after our manipulation of basis choices and field redefinition, the two effective actions \eqref{EFTaction2} and \eqref{ceaction} appear nearly identical. This is not surprising since they ultimately describe the same physical phenomenon of loop effects. However, the main difference is that, in action \eqref{ceaction}, the coefficient $c_{2}\sim\lambda_{2}\ln(\bar{r}_{h}^{2}\Lambda^{2})$ explicitly contains a black hole parameter $\bar{r}_h$ and is simply treated as a constant in that formalism. Thus, it is only applicable for extracting information from a fixed black hole metric, i.e., the extremal RN case in the present context. However, for any issues involving variations away from the given black hole metric (for example, a self-consistent analysis of the first law of thermodynamics linking neighboring black hole solutions), this approach can easily give rise to conceptual problems when interpreting the results.

Firstly, the second term of eq.\eqref{ceaction} are often viewed as a total-derivative term and completely overlooked \cite{Arkani-Hamed:2021ajd,Bittar:2024xuc}. If doing this, one can still obtain the required super-extremality of WGC (controlled by $\lambda_1$) but would not find the other side of the story, namely, the logarithmic corrections to the black hole entropy (controlled by $\lambda_2$). Secondly, one might explain the absence of $\lambda_2$ in eq.\eqref{smarrreg} by arguing that the second term of eq.\eqref{ceaction} is topological and does not affect the equations of motion \cite{Charles:2019qqt}. However, since the entropy contains a term $\lambda_{2}\ln(\bar{r}_{h}^{2}\mu_{*}^{2})$ for both the non-extremal and extremal cases, the fact is that it is essential for the black hole mass to have a corresponding term $\sim \lambda_{2}\frac{1}{\bar{r}_{h}}$ to maintain the balance (otherwise, the thermodynamic law $dM = TdS+\Phi dq$ cannot hold). This term $ \lambda_{2}\frac{1}{\bar{r}_{h}}$ is not in a logarithmic form, so we did not write it explicitly in eqs.\eqref{massnonE} and \eqref{smarrreg}, but actually there it is when examining the complete formula. So one can't simply omit the second term of eq.\eqref{ceaction} by arguing it is topological. This indicates that the analysis should be more meticulous than initially anticipated.

We now provide a consistent explanation from the perspective of our effective action  \eqref{EFTaction2} for why $\lambda_2$ appears in the logarithmic term of black hole entropy, while it doesn't appear in the logarithmic term of the mass. The second term of eq.\eqref{EFTaction2} can be decomposed into two parts: $\lambda_2\ln{\frac{\Box}{\mu_*^2}}=\lambda_2 \ln{\Box}- \lambda_2 \ln{(\mu_*^2)}$. Then, in eq.\eqref{EFTaction2}, the terms  proportional to $\lambda_2\ln(\mu_{*}^{2})$ indeed form the Gauss-Bonnet term. It is well-known that a Gauss-Bonnet term can cause a constant shift in the entropy, so it results in an entropy correction of the form $\sim \lambda_2\ln(\mu_{*}^{2})$. Due to a matching mechanism between $\ln(\bar{r}_{h}^{2})$ and $\ln(\mu_{*}^{2})$, the entropy actually receives a correction of the form $\sim \lambda_2\ln(\bar{r}_{h}^{2}\mu_{*}^{2})$, where $\ln(\bar{r}_{h}^{2})$ stems from the contribution of $\ln\square$.  Regarding the expression of mass, the terms proportional to $\lambda_2\ln(\mu_{*}^{2})$ are topological and thus do not contribute to its corrections. Then the matching mechanism also prevents the presence of the term $\lambda_2 \ln{(r_h^2)} $. This explains why $\lambda_2$ doesn't present in the logarithmic term of the mass. However, the terms associated with $\lambda_2 \ln\Box$ is not truly a Gauss-Bonnet term, nothing can preclude them from modifying the metric so that it causes the mass to obtain a non-logarithmic correction of the form $\sim \lambda_{2}\frac{1}{\bar{r}_{h}}$. Such corrections happen to ensure that the thermodynamic law $dM=TdS+\Phi dq$ holds. See Ref.\cite{Xiao:2021zly} for a detailed analysis of the variation $\delta \ln \Box$ and its impact on the field equations and black hole solutions.

\end{document}